# Observing Quantum Systems

Gerhard Grössing,
*Austrian Institute for Nonlinear Studies*,
Parkgasse 9, A-1030 Vienna, Austria
e-mail: ains@chello.at

**Abstract.** An introduction to some basic ideas of the author's "quantum cybernetics" is given, which depicts waves and "particles" as mutually dependent system components, thus defining "organizationally closed systems" characterized by a fundamental circular causality. According to this, a new derivation of quantum theory's most fundamental equation, the Schrödinger equation, is presented. Finally, it is shown that quantum systems can be described by what Heinz von Foerster has called "nontrivial machines", whereas the corresponding classical counterparts turn out to behave as "trivial machines".

## 0. Introduction

Readers of this journal are well aware that the term "cybernetics", originally coined by Norbert Wiener [Wiener, 1948], derives from the ancient Greek "kybernetes", steersman. It was Heinz von Foerster who proposed to publish a series of conference proceedings devoted to "circularly-causal feedback mechanisms in biological and social systems" with the title "Cybernetics", and in the end the famous Macy Conference Proceedings of the 1940ies and –50ies were actually published under this title. (For a new edition, see [Pias, 2003]. See also, for example, [Grössing et al., 2001].)

Cybernetics, in the broadest meaning of the term (as it was widely used, for example, also by Jean Piaget), simply means the study of systems, which are characterized by some kind of circular causality. One of the first scientists to concretely develop models of circular causality, i.e., in the 1940ies, was the neuro-physiologist Warren McCulloch. In one of his later papers, he gave a pronounced statement in this regard:

> "When we seek a mechanistic explanation of the various rhythms of any and all living systems, we eventually come to conceive of their activities in terms of circularities in the nexus of their causation."  [McCulloch, 1969]

This is interesting also for a quantum physicist, essentially because any and all quantum systems must be described by typical frequencies ("rhythms"), and any attempts at explaining quantum phenomena in terms of some linear causality have remained unsatisfactory. So, throughout the last two decades, I have, time and again, returned to this set of questions and preliminary answers, which I circumscribe with the term "quantum cybernetics": the possibility to model quantum systems as circularly causal ones. (See, for example, [Grössing, 2000].) I shall give here an introduction to the "quantum problem", propose an approach in the sense just mentioned, and finally relate the latter to Heinz von Foerster's concept of a "non-trivial machine".

## 1. The wave-particle dilemma

Since its beginnings, the discussion on the foundations of quantum theory has again and again returned to one of physics' most beautiful and elegant experiments: the diffraction at a double slit, originally performed with light by Thomas Young in the year of 1801, and repeated over and over again, also with other "quanta" such as electrons, neutrons, or even C-60 molecules recently. Fig. 1 shows a very impressive example of such an experiment, this time performed with electrons. [Tonomura, 1995]

In the double-slit experiment, light ("photons"), or some ensemble of electrons, neutrons, etc., is being emitted by a source, passes the two holes of the double slit, and is finally registered at a screen. Adjusting the source such that only one quantum is emitted per chosen time step, one obtains single "dots" on the screen, or "clicks" in some electronic register, respectively. With passing time, and more and more "dots" appearing on the screen, a distribution pattern is registered as in Fig. 1 a – d.

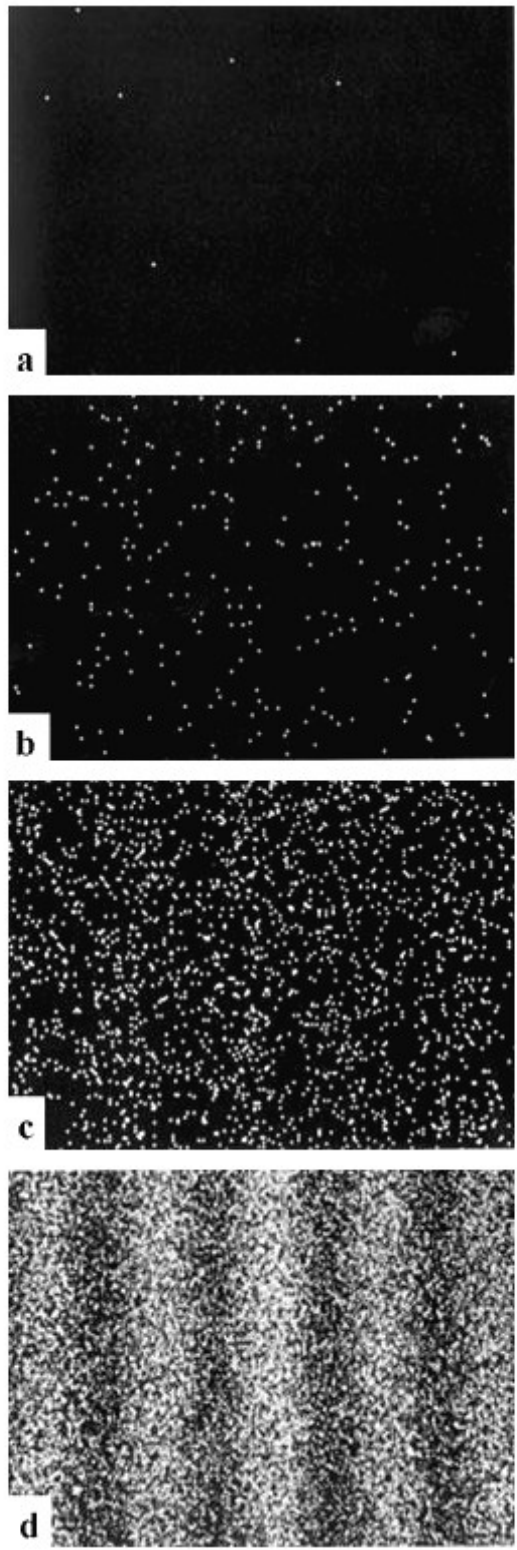

**Fig. 1 a - d**: Interference pattern produced by a) N = 8, b) N = 100, c) N = 3000, and d) N = 100 000 electrons. [Tonomura, 1995]

So, one recognizes at first, that a discrete dot is being registered per observation event, which leads to the conclusion that light (and also electrons, neutrons, etc.) must be particle-like. However, if one waits long enough (Fig. 1 c, d), it turns out that the registered particles are not distributed as one would expect if the quanta were small balls being shot through either of the two slits. Rather, they are distributed as if they interfered in a wave-like manner to produce the observed interference fringes on the screen.

In fact, if only one of the two slits, A or B, were open and the other one closed, one would obtain a distribution pattern just as if the quanta were little balls or projectiles passing through the single slit. So, here's the "magic": How does a single particle "know", when passing one slit (say, A), whether the other slit (B) was open or closed? Somehow the particle has to "know", because it would land on different regions of the screen, depending on whether or not the second slit was open. This is, in Richard Feynman's words, the "only mystery" of quantum theory, although it may be added that the mystery deepens when considering the experimentally established fact of the "nonlocality" of quanta: A particle, under specific experimental arrangements, may "know" about the goings-on in regions of space which may be so far away that not even information conveyed with the speed of light could have reached it – and still there must be ways to "in-form" the nonlocally "entangled" particles beyond what is generally assumed to be allowed from Special Relativity (i.e., that no particles can travel faster than with the speed of light).

The prevailing attitude of a majority of quantum physicists for many decades has been that this "mystery" cannot be understood and that one would have to solely rely on the exact quantum mechanical formalism, which would provide the right answers for all practical purposes anyway. However, practically since the beginnings of quantum theory in the 1920ies, there have been put forward attempts at understanding the behavior of quantum systems, now usually called the deBroglie-Bohm interpretation (dBBi). In opposition to the dominant, so-called Kopenhagen interpretation, which maintains that no causal understanding of the events in the quantum regime was possible, adherents of the dBBi were able to show that an exact causal imaging (via so-called "hidden parameters", attributed to some, yet unknown, sub-quantum medium) is in fact possible, provided that the causality is a "contextual" (or "systemic", or "holistic") one and that quantum systems are inherently nonlocal.

In fact, it was the dBBi that has pushed the topic of nonlocality into today's focus of general interest in quantum physics, because the orthodox (Kopenhagen) physicists were for decades reluctant to face the operational concreteness of the phenomenon, i.e. a quality beyond mere formalism. The one physicist responsible for this general change in attitude was John Bell, who conclusively showed in the 1960ies that the only viable "hidden variable theories" *had* to be nonlocal ones (i.e., like the dBBi), and that quantum theory must be regarded as describing nonlocal phenomena. (Curiously, many defenders of the Kopenhagen position have decided – out of ignorance and/or ideology-based malevolence - to proclaim that Bell had refuted *all* hidden variable theories, which is simply false, but which had nevertheless had a big impact on the acceptance and funding of research proposals ever since.) Regarding his attitude towards the dBBi, Bell wrote:

> "I have always felt ... that people who have not grasped the ideas of those [i.e., Bohm's 1952] papers . . . and unfortunately they remain the majority . . . are handicapped in any discussion of the meaning of quantum mechanics." [Bell, 1987]

Unfortunately, the majority of quantum physicists even today does not get involved in thinking in terms of the dBBi, which is in large parts due to the fact that the existing quantum mechanical formalism a) works perfectly, and b) has always been sufficient for new technological advances, such that the so-called "philosophical issues" on the ontology of quanta are usually put aside. However, let us remember what John Bell said with regard to the dBBi account of the double-slit experiment:

"Is it not clear from the smallness of the scintillation on the screen that we have to do with a particle? And is it not clear, from the diffraction and interference patterns, that the motion of the particle is directed by a wave? De Broglie showed in detail how the motion of a particle, passing through just one of two holes ... could be influenced by waves propagating through both holes. And so influenced that the particle does not go where the waves cancel out, but is attracted to where they cooperate. This idea seems to me so natural and simple, to resolve the wave-particle dilemma in such a clear and ordinary way, that it is a great mystery to me that it was so generally ignored." [Bell, 1987]

As can be seen from this quotation, in the dBBi one assumes that a "particle" trajectory exists which is accompanied by the propagation of waves. This is clearly reminiscent of the fact that classical mechanics can be considered as the geometrical-optical limiting case of a wave movement [Goldstein, 1953]: light (or other) rays orthogonal to wave fronts correspond to particle trajectories orthogonal to surfaces with constant "action function" $S$, where

$$S(\mathbf{x},\mathbf{p},t) = W(\mathbf{x},\mathbf{p}) - Et, \qquad (1.1)$$

with $\mathbf{x},\mathbf{p},t$ denoting location, momentum, and time coordinates, respectively, $E$ the energy, and $W$ the time-independent so-called "characteristic function". (Fig. 2)

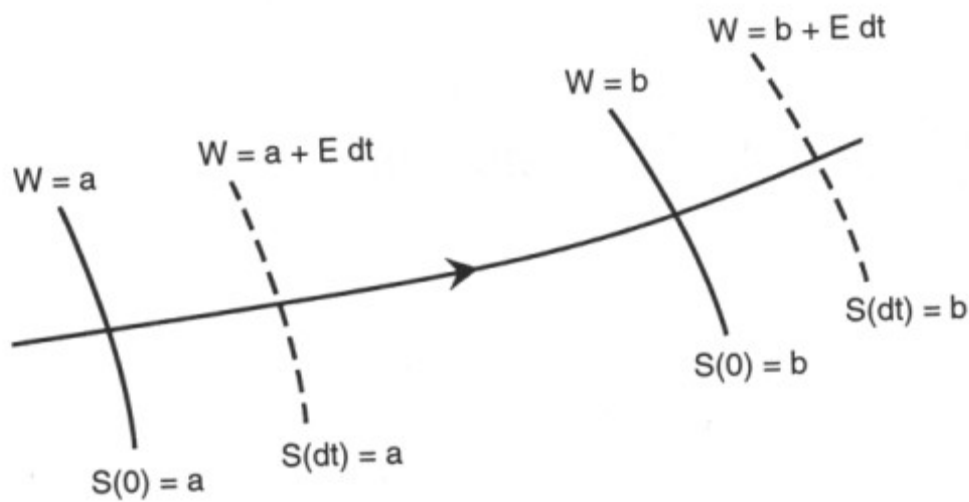

Fig. 2: Surfaces of constant action function $S$ representing wave fronts, with orthogonal particle trajectory.

With the "nabla" operator as usual providing the gradient of the operand, i.e., $\nabla \equiv \left(\frac{\partial}{\partial x}, \frac{\partial}{\partial y}, \frac{\partial}{\partial z}\right)$, the velocity $u$ of the wave fronts can be calculated from the constancy of $S$, i.e.,

$$dS = \nabla S \cdot d\mathbf{x} + \left(\frac{\partial S}{\partial t}\right)dt = |\nabla S|\mathbf{n} \cdot d\mathbf{x} + \left(\frac{\partial S}{\partial t}\right)dt = 0, \quad (1.2)$$

where

$$\mathbf{n} = \nabla S / |\nabla S| \quad (1.3)$$

is a unit vector perpendicular to the surface $S = \text{constant}$ at point $\mathbf{x}$. As $\mathbf{n} \cdot d\mathbf{x} = ds$, the component of $d\mathbf{x}$ along the surface-normal, at time $t$ the speed of the wave-front at point $\mathbf{x}$ is given by

$$u(\mathbf{x},t) = \frac{ds}{dt} = -(\partial S/\partial t)/|\nabla S|. \quad (1.4)$$

Note that $u$ mostly does not coincide with the particle velocity, but can even be much larger than the vacuum speed of light $c$ as, for instance, when $(-\partial S/\partial t) = E = mc^2$ and $\nabla S = \mathbf{p} = m\mathbf{v}$. Note also that the example of Fig. 1 is only a particular one, and that another wide-spread example is given by *spherical wave surfaces* according to *Huygens' principle*. Still, in classical mechanics, said particle trajectories (or "rays") and the propagating wave-fronts are not causally related to each other, and the wave fronts are seen mostly as a curious formal feature only.

Let us now turn to quantum theory. In the usual dBBi, the particle simply responds to the local value of the wave-like field in its vicinity, but there is no reciprocal action of the particle on the wave. This is commented by Peter Holland in the following way: "From the standpoint of general theoretical principles this feature of the causal interpretation may appear as unsatisfactory, calling for a development of the theory to include a more symmetrical relation between wave and particle." [Holland, 1993, p.120] What I propose as a "quantum cybernetics", in fact, owes its name essentially to a correspondingly proposed circularly causal relationship between wave and "particle" (with the quotation marks to be explained below).

However, some care has also to be taken as to what one considers as a physically effective "wave". Quantum cybernetics agrees with orthodox positions (including the dBBi) that "particles" cannot "change" the quantum mechanical "wave function". The reason for this is that thereby one would only confuse levels of description: the wave function, in the dBBi as in all other interpretations, is a compact description of the *total knowledge* about a quantum system, with no immediate "physical counterpart". However, "particles" in quantum cybernetics can change the above-mentioned spherical Huygens waves and thus the details of *individual events* in the surrounding field (i.e., still in accordance with an *overall* state of affairs as described by the wave function). In other words, one can implement a circular causality on a supposed *sub-quantum level*. As I will show below, "particles" can thus be considered, via permanent emission of Huygens waves (i.e., surfaces of constant action function $S$) to co-determine the structure of the total surrounding wave-like field, and the latter in turn determines where the "particles" can go.

In fact, one can take as a starting point the above mentioned orthogonality in classical mechanics between wave fronts and particle trajectories. (For certain types of external potentials, this orthogonality is disturbed, but can be re-obtained with the introduction of a suitable metric. [Holland, 1982]) Erwin Schrödinger, in a talk given in 1952, has pointed out where to look for a transitional domain between classical wave mechanics and quantum mechanics:

> "At each place in a regularly propagating wave-train, one finds a two-fold structural connection of actions, which one may distinguish as 'longitudinal' and 'crosswise', respectively. The crosswise structure is the one along wave-planes; it manifests itself in diffraction and interference experiments. The longitudinal structure, in turn, is the one of the wave-normal and manifests itself when individual particles are observed. Both structural connections are completely confirmed, namely through useful experimental arrangements, each carefully designed for its particular purpose.
>
> However, the notions of 'longitudinal' and 'crosswise' structures are no precise ones, no absolute ones, just because the notions of wave-planes and wave-normals are not. These notions of wave-planes and wave-normals, and thus the differentiation between longitudinal and crosswise structures, are necessarily getting lost when the whole wave phenomenon is reduced to a small spatial region of the size of a single or only a few wavelengths. And this case now is of particular interest." [Schrödinger, 1952]

However, we shall see shortly that said differentiation is not completely getting lost, essentially because a "wave phenomenon" in general cannot be reduced to a small region just by considering the "main bulk" of a wave packet only. In fact, it has been demonstrated conclusively in so-called "quantum post-selection experiments" [Werner 1991, Jacobson, 1994] that the handling of small-sized "wave packets" (to which Schrödinger's comment alludes) does not exclude the possibility to talk about wave-planes and particle trajectories: Interference between two possible quantum paths can be demonstrated, although the two corresponding bulks of the wave packets may never overlap (i.e., in $\mathbf{x}$-space). Rather, it is the non-locally extending plane-wave components of the wave-packets (which were often thought of as negligible), which make interference possible (i.e., in $\mathbf{k}$-space). In general, of course, the relation between wave-planes and particle paths will be more complicated than the simple orthogonality in classical mechanics.

Summarizing, one can say that the transition from classical to quantum mechanics is characterized by a transition from a general orthogonality between wave-fronts and particle trajectories, respectively, to a situation where orthogonality is an exception rather than the rule, the latter being a non-orthogonal relationship between particle trajectories and waves. In the next chapter, we shall consider a derivation of the Schrödinger (and the Klein-Gordon) equation under exactly this premise, i.e., thereby starting off with some basic equations of classical mechanics.

## 2. Derivation of the Schrödinger and Klein-Gordon equations from classical mechanics

In the following, I want to take a look back onto the early twentieth century and delineate a then possible line of developments towards the establishment of quantum theory, which is different from the line that actually did turn out. In doing so, however, we shall see that even nowadays one could have a very different, but still also very useful view of the quantum world (thereby "enhancing the number of possibilities for action", as Heinz von Foerster would say).

It was established in 1905 by Albert Einstein, through his correct description of the "photoelectric effect" (which brought him the Nobel price), that light "particles" ("photons") of frequency $\omega$ exchange with their environment only discretized chunks of energy such that the energy of a "particle" can be written as

$$E = \hbar\omega, \tag{2.1}$$

with $\hbar$ being Planck's constant $h$ divided by $2\pi$. In 1927, then, Louis de Broglie showed that any "particle" has also wave-like attributes such that its momentum **p** can be written in terms of the associated wave-number **k** (i.e., proportional to the inverse of the wave-length $\lambda$),

$$\mathbf{p} = \hbar\mathbf{k}. \tag{2.2}$$

With the introduction of four-vectors in Einstein's Special Theory of Relativity (i.e., one time component, and three components in space), equations (2.1) and (2.2) can be written in the compact form

$$p^\mu = \left(\frac{E}{c}, \mathbf{p}\right) = \hbar k^\mu = \hbar\left(\frac{\omega}{c}, \mathbf{k}\right), \tag{2.3}$$

thus introducing the energy $E$ as the (time-related) zero-component of the four-momentum $p^\mu$. In what follows, I shall use only expression (2.3) and the drawing on the essentials of chapter 1 of this paper, i.e., the orthogonality of wave- and particle-related vectors in classical mechanics, to derive quantum mechanics by showing how this orthogonality is generally lost, and under which circumstances this is so.

At this point a remark is necessary as to what is here understood as a "derivation". Of course, the Schrödinger equation has been known for a long time by now, and any attempt to "derive" it may be suspicious of forcing things towards an outcome that is already well known. However, as I will try to show, what is proposed as a "derivation" here deserves its name due to the fact that the premises are more encompassing than what is usually regarded as "classical physics". In fact, the essential ingredient which makes it possible to "derive" quantum theory from "classical physics" is given by *explicitly taking as physical the Huygens waves of the Hamilton-Jacobi theory*. In other words, I derive the Schrödinger equation from classical physics with the *additional assumption* that the dynamics of Huygens waves are *physically effective* (as opposed to a mere formal meaning in the "classical" Hamilton-Jacobi theory).

The suggested procedure will be the following. Firstly, we perform a usual ("virtual") variation on a classical Lagrangian. Then, we shall put the resulting continuity equation into a form that allows us to glean general expressions for fluctuating energy and momentum components, respectively. Assumed to now represent "real" (i.e., physically effective, and indispensable) variations, the latter will then be re-introduced as additional elements into a correspondingly modified Lagrangian. Renewed variation of the thus obtained Lagrangian will then provide the Schrödinger equation. This specific type of "second order" variational procedure will then, towards the end of the paper, be found as essential for a comparison with the dynamics of quantum systems on an operational level: quantum systems can be shown to operate like "non-trivial machines", as opposed to the "trivial machines" of classical Hamiltonian systems.

As is usual in the standard Hamilton-Jacobi formulation of classical mechanics, we start with a formulation of the action integral for a single free particle in an $n$-dimensional configuration space. With unknown initial position (which itself is due to some hypothesized, yet unknown, sub-quantum dynamics), the latter must be described via a probability density distribution. With the four-vector notation $dx^\mu := (cdt, d\mathbf{x})$, the usual sum convention, and in accordance with relativistic kinematics, the action integral can be formulated with the Lagrangian $L$ (see, for example, [Holland, 1993], or [Hall, 2002]) as

$$\mathrm{A} = \int L d^n x dt = \int P \left\{ \frac{1}{m} \partial_\mu S \partial^\mu S + \frac{\partial S}{\partial t} \right\} d^n x dt = \int P \left\{ \frac{1}{m} p_\mu p^\mu + \frac{\partial S}{\partial t} \right\} d^n x dt. \quad (2.4)$$

A non-relativistic variant thereof (with the introduction of some external potential $V$) reads

$$\mathrm{A} = \int L d^n x dt = \int P \left\{ \frac{1}{2m} \nabla S \cdot \nabla S + \frac{\partial S}{\partial t} + V \right\} d^n x dt. \quad (2.5)$$

Here the action function $S(\mathbf{x}, t)$ is related to the particle velocity $v^\mu = p^\mu / m$ via

$$v^\mu = \frac{1}{m} \partial^\mu S, \quad (2.6)$$

or, non-relativistically,

$$\mathbf{v} = \frac{1}{m} \nabla S, \quad (2.7)$$

with the usual four-derivative $\partial^\mu := \left( \frac{1}{c} \frac{\partial}{\partial t}, \nabla \right)$, and the probability density $P(\mathbf{x}, t)$ that a particle is found in a given volume of configuration space being normalized such that

$$\int P d^n x = 1. \quad (2.8)$$

Let us now perform a "virtual variation" of the Lagrangian in equation (2.4) and consider its result. Upon fixed end-point variation, i.e., $\delta S = 0$ at the boundaries, one has

$$\frac{\partial L}{\partial S} - \partial_\mu \left\{ \frac{\partial L}{\partial (\partial_\mu S)} \right\} = 0, \tag{2.9}$$

and thus one obtains the relativistically invariant "continuity equation" for the conservation of the probability current $J^\mu = P \partial^\mu S$ of some probability distribution $P$, i.e.,

$$\partial_\mu J^\mu = \partial_\mu \left[ P \partial^\mu S \right] = \partial_\mu P \partial^\mu S + P \partial_\mu \partial^\mu S = 0. \tag{2.10}$$

Inserting the expression for the four-momentum

$$p^\mu = m \, \mathrm{v}^\mu = \partial^\mu S \tag{2.11}$$

into equation (2.10) then provides that the four-gradient of $p^\mu$ is given by

$$\partial_\mu p^\mu = -\frac{\partial_\mu P}{P} p^\mu. \tag{2.12}$$

Thus, the conservation of the probability current $J^\mu$ entails the conservation of energy-momentum, $\partial_\mu p^\mu = 0$, *only* when the scalar product on the r.h.s. of (2.12) vanishes, i.e., when the two vectors are orthogonal, i.e., $\frac{\partial_\mu P}{P} p^\mu = 0.$

Equation (2.12) will be the starting point on our way to derive two basic equations of quantum theory, i.e., the Schrödinger equation and its relativistic generalisation for spinless particles, i.e., the Klein-Gordon equation. Let us first turn our attention to the nonrelativistic regime, i.e., where $E/c^2 \to 0$. The corresponding version of equation (2.12) reads

$$\nabla \cdot \mathbf{p} = -\frac{\nabla P}{P} \cdot \mathbf{p}. \tag{2.13}$$

With $\mathbf{p} = m\mathbf{v}$ we observe that the classical, so-called "Hamiltonian flow" (i.e., of incompressible fluids) given by

$$\nabla \cdot \mathbf{v} = 0 \tag{2.14}$$

is only obtained if the r.h.s. of (2.13) vanishes, too, i.e.,

$$\left( \frac{\nabla P}{P} \right) \cdot \mathbf{v} = 0. \tag{2.15}$$

Moreover, with equation (2.2) we can also rewrite (2.13) as

$$\nabla \cdot \mathbf{k} = -\frac{\nabla P}{P} \cdot \mathbf{k}, \tag{2.16}$$

with classical Hamiltonian flow then implying that the vectors $\frac{\nabla P}{P}$ and $\mathbf{k}$ are orthogonal.

In general, equation (2.13) is an expression for the non-conservation of momentum $\mathbf{p} = m\mathbf{v}$. Upon the "virtual" variation of the action function $S$ we thus arrive at an expression for a "real" (i.e., physically effective) variation $\delta S$, which does exist as a physical possibility once the "classical" orthogonality condition is discarded, and which involves some additional momentum $\delta \mathbf{p}$. This is due to the fact that we have to do with a probability field $P$, which in turn refers to the assumed sub-quantum stochastic processes, and which can manifest itself via possible momentum fluctuations $\delta \mathbf{p}$.

Thus, just as in equation (1.2) for wave-planes with $S = $ constant, we can now for the fluctuating terms consider additional wave-planes with $\delta S = $ constant, i.e.,

$$d(\delta S) = \nabla(\delta S) \cdot d\mathbf{x} + \left(\frac{\partial(\delta S)}{\partial t}\right) dt = |\nabla(\delta S)| \mathbf{n} \cdot d\mathbf{x} + \left(\frac{\partial(\delta S)}{\partial t}\right) dt = 0, \tag{2.17}$$

where

$$\mathbf{n} = \nabla(\delta S) / |\nabla(\delta S)| \tag{2.18}$$

is a unit vector perpendicular to the surfaces of $\delta S = $ constant at point $\mathbf{x}$.
As according to equation (2.13) it holds that

$$\frac{\nabla P}{P} \cdot d\mathbf{x} + (\nabla \cdot \mathbf{v}) dt = 0, \tag{2.19}$$

and remembering the factor $\hbar$ in equs. (2.1) and (2.2), which of course is also relevant for any *fluctuations* in energy or momentum, respectively, we can write

$$d(\delta S) = |\nabla(\delta S)| \mathbf{n} \cdot d\mathbf{x} + \left(\frac{\partial(\delta S)}{\partial t}\right) dt = 0 = \alpha \hbar \left\{ \left|\frac{\nabla P}{P}\right| \mathbf{n} \cdot d\mathbf{x} + (\nabla \cdot \mathbf{v}) dt \right\}, \tag{2.20}$$

where $\alpha$ is some proportionality constant.

To fix the latter, we can either use a theoretical argument, or refer to experimental evidence: both approaches will lead to the same value. First, we can assume that, as the intensity $P$ of a wave is given by the square of its amplitude $R$, i.e., $R = \sqrt{P}$, we can define a wave-number $\mathbf{k}_u$ such that its size equals simply the relative gradient of $R$, i.e.,

$$|\mathbf{k}_u| := \left|\frac{\nabla R}{R}\right| = \frac{1}{2}\left|\frac{\nabla P}{P}\right|, \tag{2.21}$$

and therefore, comparing with equation (2.20), that

$$\alpha = \frac{1}{2}. \tag{2.22}$$

Returning now to the intention formulated at the beginning, i.e., to have a look back in time and see how things might have developed differently, one could also start with an experimental finding from the year 1924 [Mulliken, 1924]: It was then found that to each energy $E = \hbar\omega$ of a particle, one must associate (in three spatial dimensions) an additional term of a "zero-point energy"

$$E_0 = 3\frac{\hbar\omega}{2}, \tag{2.23}$$

which refers to a fluctuating field in the vacuum. Now, with the ansatz (which will be detailed in the last chapter) that

$$-\frac{\partial(\delta S)}{\partial t} = \delta E = \frac{3}{2}\hbar\omega = \frac{1}{2}\hbar\left(\nabla \cdot \left[\frac{1}{\delta t}\delta\mathbf{x}\right]\right) = \frac{1}{2}\hbar(\nabla \cdot \mathbf{v}), \tag{2.24}$$

we obtain again that in equation (2.20), $\alpha = 1/2$.

So, we arrive at the following physical picture for classical Hamiltonian flow. A particle has momentum $\mathbf{p} = \hbar\mathbf{k}$, and from its actual position there are permanently spherical waves emitted with wave-number $\mathbf{k}_u$, such that the wave-planes with action function $S = \text{constant}$ and the particle trajectories along $\nabla S$ are orthogonal.

However, let us now consider the more interesting situation where this orthogonality is disturbed. In other words, we'll have to consider a *contextual* situation, where not only the waves originating from the particle position are relevant, but where said waves interfere or are in some other way related to other waves from the surrounding environment. In other words, we shall put *constraints* on our system, which will have an effect on the probability distribution $P$ across the whole volume which is relevant for the system's dynamics. And, as $P$ will be affected, so will be our wave number $\mathbf{k}_u$, according to equation (2.21). The non-orthogonality of $\mathbf{k}_u$ and $\mathbf{k}$ then implies that the momentum $\mathbf{p} = \hbar\mathbf{k}$ in general is not a conserved quantity any more. Rather, momentum fluctuations have to be taken into account, which means that the direction of the vector $\mathbf{k}_u$ may change spontaneously and uncontrollably. In other words, whenever $\mathbf{k}_u$ is not orthogonal to $\mathbf{k}$, but instead allowed to fluctuate randomly (i.e., either due to "internal" spontaneous action, or because of a contextual shift in the $P$-field), the momentum $\mathbf{k}$ itself will also undergo a fluctuation, a fact which will be shown below to lead from "classical" to "quantum" motion. (cf. Fig. 3)

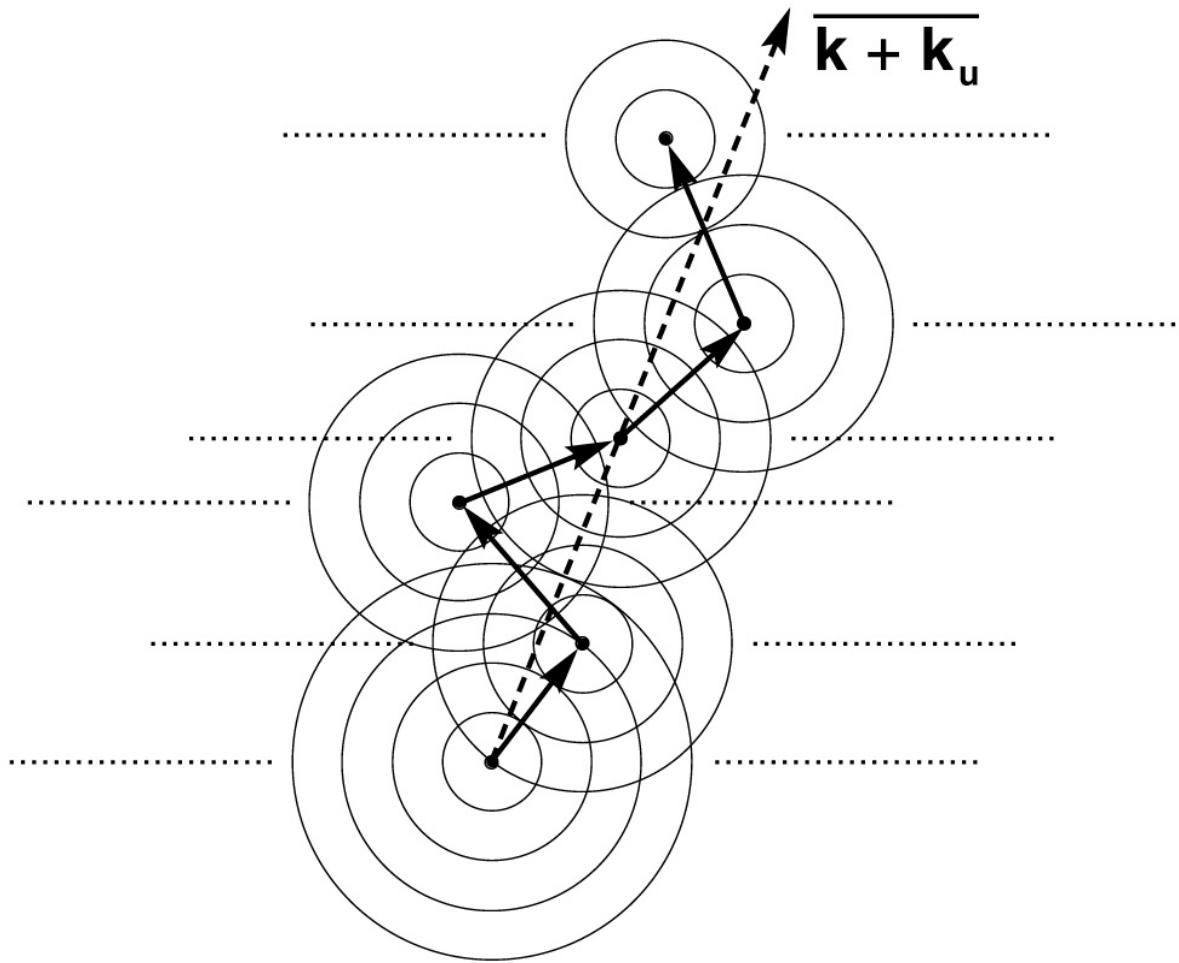

Fig. 3: Schematic representation of quantum flow including fluctuations $\mathbf{k}_u$ of classical wave numbers $\mathbf{k}$, with the circles indicating the propagation of spherical Hamilton-Jacobi wave surfaces. The dotted lines indicate symbolically that the waves pictured represent only the local surroundings of a generally extending probability field $P$, thus illustrating that the fluctuations are to be seen in the context of the whole embedding environment. Note that the uncontrollability of the directions $\mathbf{n} = \mathbf{k}_u / |\mathbf{k}_u|$ reflects two possible causal mechanisms: either, $\mathbf{n}$ represents a spontaneous ("internal") fluctuation of the particle (i.e., due to some collision with a particle of the hypothesized sub-quantum aether), or it is due to the configuration of the total, non-locally connected, $P$-field, which may determine an additional momentum component of the particle. In other words, the particle can either actively co-determine (via emission of Huygens waves) the configuration of the $P$-field, or the latter may determine the particle's path by "guiding" it. This is why one can speak of a "circular causality", or of a "quantum cybernetics", respectively.

We are thus forced to conclude that $\mathbf{p} = \nabla S$ can only be an average momentum which is subject to momentum fluctuations $\delta\mathbf{p}$ around $\nabla S$. Thus, the physical momentum is given by

$$\mathbf{p} := \nabla S + \delta\mathbf{p}. \tag{2.25}$$

We shall assume with [Hall, 2002] that the *average* momentum fluctuations, denoted $\overline{\delta\mathbf{p}}$, are linearly uncorrelated with the average momentum $\overline{\nabla S}$. In other words, with the fluctuations themselves being considered unbiased, i.e., $\int P\overline{\delta\mathbf{p}}d^n x = 0$, the average over fluctuations *and* position of the product of the two momentum components are also assumed to be unbiased:

$$\int P\overline{(\nabla S \cdot \delta\mathbf{p})}d^n x = 0. \tag{2.26}$$

Thus, it is proposed that the new action integral of our particle is given by

$$A := \int P \left\{ \frac{\partial S}{\partial t} + \frac{\overline{\mathbf{p}} \cdot \overline{\mathbf{p}}}{2m} + V \right\} d^n x dt, \tag{2.27}$$

with $\overline{\mathbf{p}} = \overline{\nabla S + \delta\mathbf{p}}$. With condition (2.26), this provides

$$A = \int P \left\{ \frac{\partial S}{\partial t} + \frac{1}{2m}\nabla S \cdot \nabla S + V \right\} d^n x dt + \frac{1}{2m}\int \left(\Delta\overline{p}\right)^2 dt, \tag{2.28}$$

where $\Delta\overline{p}$ is the average rms momentum fluctuation, i.e.,

$$\left(\Delta\overline{p}\right)^2 = \int P\overline{\delta\mathbf{p}\cdot\delta\mathbf{p}}d^n x = \int P|\delta\mathbf{p}|^2 d^n x. \tag{2.29}$$

Now, although we cannot know the direction of the assumed fluctuation, $\mathbf{n} = \delta\mathbf{p}/|\delta\mathbf{p}|$, which is subject to some (yet unknown) sub-quantum stochastic process, we actually know the size $|\delta\mathbf{p}|$ from equations (2.20) or (2.21), respectively, such that

$$|\delta\mathbf{p}| = \hbar|\mathbf{k}_u| = \frac{\hbar}{2}\left|\frac{\nabla P}{P}\right|. \tag{2.30}$$

In fact, the uncontrollability of the directions $\mathbf{n} = \mathbf{k}_u/|\mathbf{k}_u|$, symbolized by the "erratic" fluctuations of the small vectors of variable length in Fig. 3, reflects two possible causal mechanisms: either, $\mathbf{n}$ represents a spontaneous ("internal") fluctuation of the particle (i.e., due to some collision with a particle of the hypothesized sub-quantum aether), or it is due to the configuration of the total, non-locally connected, $P$ - field, which may determine an additional momentum component of the particle. In other words, the particle can either actively co-determine (via emission of Huygens waves) the configuration of the $P$ - field, or the latter may determine the particle's path by "guiding" it.

Inserting (2.30) into (2.29), we get

$$\left(\overline{\Delta p}\right)^2 = \int P \left(\frac{\hbar}{2}\left|\frac{\nabla P}{P}\right|\right)^2 d^n x. \qquad (2.31)$$

Thus, combining equations (2.28) and (2.31), we obtain the new (and now complete) action integral for "contextual" particles as

$$A := \int P \left\{ \frac{\partial S}{\partial t} + \frac{(\nabla S)^2}{2m} + \frac{\hbar^2}{8m}\left(\frac{\nabla P}{P}\right)^2 + V \right\} d^n x \, dt. \qquad (2.32)$$

Now, upon renewed fixed end-point variation in $S$ nothing changes, i.e., we again obtain the continuity equation (2.10). However, performing now the fixed end-point variation in $P$, i.e.,

$$\frac{\partial L}{\partial P} - \frac{\partial}{\partial x_i}\left\{\frac{\partial L}{\partial\left(\frac{\partial P}{\partial x_i}\right)}\right\} = 0, \qquad (2.33)$$

one obtains the so-called Hamilton-Jacobi-Bohm equation, i.e.,

$$\frac{\partial S}{\partial t} + \frac{(\nabla S)^2}{2m} + V + \frac{\hbar^2}{4m}\left[\frac{1}{2}\left(\frac{\nabla P}{P}\right)^2 - \frac{\nabla^2 P}{P}\right] = 0. \qquad (2.34)$$

The last term on the l.h.s. of equation (2.34) in the usual dBBi is called the "quantum potential", but I want to stress here that this non-classical term in the context presented here is due to a kinetic energy term (viz., equations (2.31) and (2.32)) rather than a potential energy, as was similarly argued previously by [Harvey, 1966]. Still, as is well known [Bohm, 1952], the equations (2.10) and (2.34), together with the introduction of the complex-numbered "wave function"

$$\psi = \sqrt{P} \exp\{-iS/\hbar\}, \qquad (2.35)$$

can be condensed into a single equation, i.e., the Schrödinger equation

$$i\hbar \frac{\partial \psi}{\partial t} = \left(-\frac{\hbar^2}{2m}\nabla^2 + V\right)\psi. \qquad (2.36)$$

Extension to a many-particle system is straightforwardly achieved by starting the same procedure with a correspondingly altered Lagrangian in (2.32), which then ultimately provides the usual many-particle Schrödinger equation.

Finally, we can now also return to our fully relativistic equation (2.4). From (2.12) we see that our momentum fluctuation term must now be

$$\left|\hbar k_u^{\mu}\right| = \left|\frac{\hbar}{2}\frac{\partial^{\mu}P}{P}\right|, \tag{2.37}$$

which again leads to a correspondingly modified action integral, i.e.,

$$A = \int L d^n x dt = \int P\left\{\frac{1}{m}\partial_{\mu}S\partial^{\mu}S + \frac{\hbar^2}{4m}\frac{\partial_{\mu}P}{P}\frac{\partial^{\mu}P}{P} + \frac{\partial S}{\partial t}\right\}d^n x dt, \tag{2.38}$$

finally providing (with the usual quabla operator $\Box := \partial_{\mu}\partial^{\mu}$)

$$\partial_{\mu}S\partial^{\mu}S := M^2 c^2 = m^2 c^2 + \hbar^2 \frac{\Box\sqrt{P}}{\sqrt{P}}. \tag{2.39}$$

Again, as is well known, equations (2.10) and (2.39) can be written in compact form by using the "wave function" $\Psi = \sqrt{P}\exp(-iS/\hbar)$ to obtain the usual Klein-Gordon equation

$$\left(\Box + \frac{m^2 c^2}{\hbar^2}\right)\Psi = 0. \tag{2.40}$$

We have thus succeeded in deriving the Schrödinger- and Klein-Gordon equations from classical Lagrangians with a minimal set of additional assumptions relating to an additional momentum fluctuation (viz., equations (2.30) or (2.37), respectively) associated to each particle. For a more detailed discussion of this new derivation, including also a derivation of Heisenberg's uncertainty relations, see [Grössing, 2004].

### 3. Quantum Cybernetics

In chapters 2 and 3, we have seen how one can understand quantum mechanics, not only per se, but also in relation to classical mechanics. In the latter, the "context" of a "particle" as it is given by the spatial characteristics of the probability field $P$, is practically invisible or ineffective, because the mathematical representation of this context, i.e., the vector field $\mathbf{k}_u$ representing spherical waves emanating from the "particle", is always orthogonal to the momentum $\mathbf{p} = \hbar\mathbf{k}$ of the "particle" such that no extra force (apart from those due to existing classical potentials) is acting on the latter.

However, the quantum regime can actually be defined by the situation where said orthogonality disappears, thus attributing an active role to the vector $\mathbf{k}_u$, which manifests itself in a causal quantum theory of motion, characterized by features of the particle's (generally nonlocal) context.

Moreover, in this viewpoint "particles" and waves are treated on an equal footing: "particles", via permanent emission of Huygens waves, co-determine the structure of the total surrounding wave-like field (cf. Fig. 3), and the latter in turn determines the trajectories along which "particles" can go. This is the essence of what I call "quantum cybernetics": the circularly causal relation between waves and "particles" [see, e.g., Grössing, 1986 or Grössing, 2002, and the monograph Grössing, 2000]. Having read the latter in 2000, Heinz von Foerster in an interview [Grössing and von Foerster, 2000] praised my "ingenious idea" (in his typical exaltations of charming compliments), which in his view came down to saying: "There exists no point per se, because there's always an environment, too." Of course, this was not really my idea, and my approach is just a new formulation embedded within a long-lasting tradition, dating back at least to Louis de Broglie's attempts at a "non-linear wave mechanics" [de Broglie, 1960]. Still, today's "elementary particles" like the electron, for example, are assumed to be point-like (though surrounded by a cloud of "virtual particles" from the quantum vacuum). However, a point is an idealized mathematical construction and always will be only that, because in order to prove its existence operationally, an infinite precision would be required, which is impossible. Thus, alternatives to the "point-like" paradigm are not only legitimate, but even mandatory, before one would have to accept that quantum theory was full of "magic" and "mysteries", as many a popular account would like to have it.

This is how the idea of quantum cybernetics came into being. Based on the dBBi of quantum theory, but not convinced by its duality between "particles" and their (guiding) wave-like environments, which only allows influences of the latter on the former, but not vice versa, I have aimed at a mutually causal relation between "particles" and waves. In asking, "How does a quantum system 'perceive' its environment?" [Grössing, 1988], I re-examined de Broglie's old idea that a "particle" may actually be the highly non-linear, soliton-like part of an elsewhere linear wave. This picture then served as a basis for the proposition that it is the oscillations of the quantum system that serve as a means to "observe" the environment. Following Francisco Varela and his definition of an "autonomous system" [Varela, 1979], I proposed [Grössing, 1988]: "A single pure-state quantum system is a feedback system with a given reference signal that compensates disturbances only relative to the reference point (i.e., a basic frequency $\omega$), and not in any way reflects the texture of the disturbance. Its behavior, then, is the process by which such a unit controls its 'perceptual data' through adjusting the reference signal." How can one envisage the corresponding "organizational closure" of a quantum system? First, note with [Maturana, 1980]: "If one says that there exists a machine $M$ in which there is a feedback loop through the environment, so that the effects of its output affect its input, one is in fact talking about a larger machine $M'$ which includes the environment and the feedback loop in its defining organization."

So, the task is to find a description involving the whole quantum system, i.e. the quantum together with its context as given by the experimenter's apparatus. De Broglie had already noted that one can consider the whole quantum system as one with a "variable rest mass", where the total energy of the system, $Mc^2 = \hbar\Omega$, decomposes into the "particle's" rest mass $m$ and the variable part $\delta m$ due to the experiment's context:

$$M^2 c^4 = \hbar^2 \Omega^2 = \partial_\mu S \partial^\mu S = \{m^2 + \delta m^2\} c^4 = \hbar^2 \{\omega^2 + \delta\omega^2\}. \tag{3.1}$$

The last term on the r.h.s. already indicates that one can in fact consider a quantum system as characterized by a typical frequency $\omega$ and its variation $\delta\omega$, which is related to the degree of non-orthogonality as discussed in chapter 2. This can be seen as follows. We recall from equation (2.13) that $\nabla \cdot \mathbf{v} = -\dfrac{\nabla P}{P} \cdot \mathbf{v},$ and thus also

$$(\nabla \cdot \mathbf{v})\,dt = -2\mathbf{k}_u \cdot d\mathbf{x}. \tag{3.2}$$

The l.h.s. of equation (3.2) can be written as

$$(\nabla \cdot \mathbf{v})\,dt = \left\{ \frac{\partial}{\partial x} \cdot \frac{1}{dt} dx + \frac{\partial}{\partial y} \cdot \frac{1}{dt} dy + \frac{\partial}{\partial z} \cdot \frac{1}{dt} dz \right\} dt. \tag{3.3}$$

If we now assume that for the system involving a "basic frequency" $\omega$, this frequency also determines the maximal temporal resolution of the system (i.e., such that time spans shorter than $1/\omega$ are not operational), we obtain a "coarse graining" of the time axis, with its inverse elementary unit becoming (with some proportionality constant $\alpha$)

$$\frac{1}{dt} \to \frac{\alpha}{\delta t} \equiv \omega. \tag{3.4}$$

Thus, $1/dt$ becomes $\omega$, and we get from (3.3) that

$$(\nabla \cdot \mathbf{v})\,\delta t = 3\omega\delta t. \tag{3.5}$$

Multiplying (3.2) with $\hbar$ and inserting (3.5) then provides

$$\hbar \mathbf{k}_u \cdot d\mathbf{x} = -3\frac{\hbar\omega}{2}\delta t. \tag{3.6}$$

Thus, we obtain

$$\hbar \mathbf{k}_u \cdot \mathbf{v} = -\frac{\hbar}{2}(\nabla \cdot \mathbf{v}) = -3\frac{\hbar\omega}{2}. \tag{3.7}$$

In other words, we have now traced back the degree of non-orthogonality to the "zero-point energy" $E_0$, which is exactly given by

$$E_0 = 3\frac{\hbar\omega}{2}. \tag{3.8}$$

As in classical Hamiltonian flow the orthogonality is always maintained, this means that then it always holds that the zero-point energy can be neglected, $E_0 \equiv 0$. Moreover, quantum theory is now seen to emerge as soon as there exists a non-orthogonality which also corresponds to a term $E_0 \neq 0$. This makes it possible now to differentiate between classical and quantum mechanics in terms of Heinz von Foerster's distinction between trivial and non-trivial machines. [von Foerster, 1991]

Whereas in classical Hamiltonian flow all expressions in (3.7) are identical to zero, the situation in the quantum domain becomes definitely nontrivial. (Fig.4)

Remembering Piaget's dictum that "recognizing reality means to construct transformational systems … which are more or less isomorphous to transformations of reality" [Piaget, 1967], we can try to capture the essentials of a quantum systems' performance in Fig. 4. Here, the nontrivial machine's "state function" is given by the (nonlocally extending) probability field $P$ which via fluctuations $\delta \mathbf{p} = \mathbf{n}\hbar |\mathbf{k}_u|$ contributes to the variable frequency part of equation (3.1), i.e., $\delta\omega$, whereas the "driving function" $M^2 \propto \Omega^2$ provides the output $\left(k^\mu = \partial^\mu S/\hbar, k_u^{\ \mu} = \partial^\mu P/2P\right)$, which via a feedback loop through the environment becomes the new input. If $\delta\omega = 0,$ so that also the terms in equation (3.7) can be put equal to zero, then the machine becomes trivial and we obtain classical Hamiltonian flow.

Finally, note that if already a single-particle quantum system as a non-trivial machine is capable of "observing" its environment, i.e., by modulation of its characteristic frequency, then quantum theory in this view is about "observing observing systems", and thus about second order cybernetics.

**Note added in proof:**

I thank the anonymous referees for their comments, and particularly one of them for the very instructive and helpful suggestions concerning the didactics of the present paper.

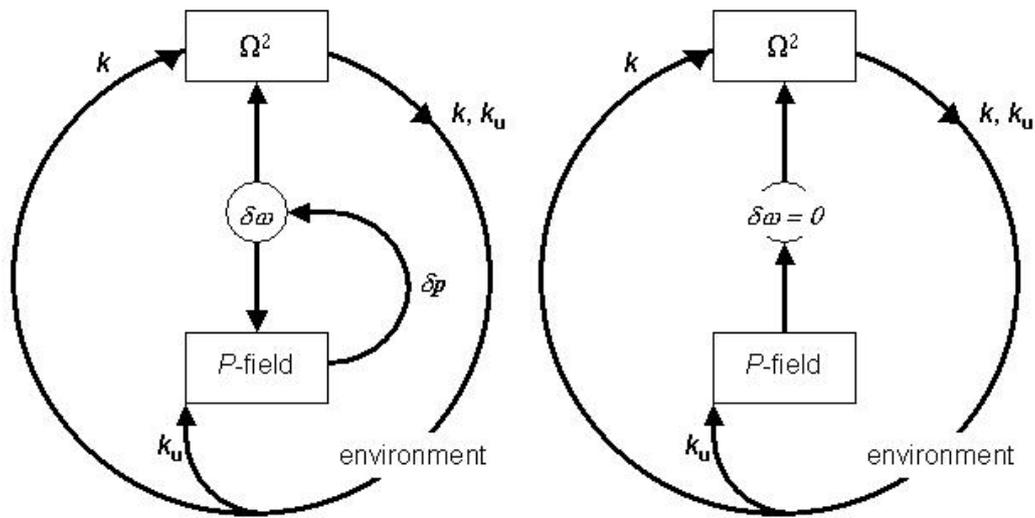

Fig. 4: Quantum system as a non-trivial machine (left) vs. classical Hamiltonian flow as a trivial machine (right). Due to the existence of "zero point fluctuations", the essential characteristic of quantum systems, as opposed to the classical counterparts, is given by momentum fluctuations $\delta \mathbf{p} = \mathbf{n}\hbar|\mathbf{k}_u|$, with uncontrollable directions $\mathbf{n}$ as shown in Fig. 3. The latter, then, can affect both the "particle" and the surrounding wave-like $P$-field.